\documentclass[conference]{IEEEtran}
\IEEEoverridecommandlockouts

\usepackage{cite}
\usepackage{amsmath,amssymb,amsfonts}
\usepackage{algorithmic}
\usepackage{graphicx}
\usepackage{textcomp}
\usepackage{xcolor}
\usepackage{balance}

\def\BibTeX{{\rm B\kern-.05em{\sc i\kern-.025em b}\kern-.08em T\kern-.1667em\lower.7ex\hbox{E}\kern-.125emX}}

\begin{document}

\title{Finite-Size CV-QKD for Satellite Quantum\\ Communication under a Hybrid Noise Model}

\author{\IEEEauthorblockN{Mouli Chakraborty\IEEEauthorrefmark{1},
Avishek Nag\IEEEauthorrefmark{2},
Subhash Chandra\IEEEauthorrefmark{3},
Anshu Mukherjee\IEEEauthorrefmark{4},\\
Konstantinos Koufos\IEEEauthorrefmark{1},
and Mehrdad Dianati\IEEEauthorrefmark{1}}
\IEEEauthorblockA{\IEEEauthorrefmark{1}School of Electronics, Electrical Engineering and Computer Science, Queen's University Belfast, UK}
\IEEEauthorblockA{\IEEEauthorrefmark{2}School of Computer Science, University College Dublin, Ireland}
\IEEEauthorblockA{\IEEEauthorrefmark{3}School of Natural Science, Trinity College Dublin, Ireland}
\IEEEauthorblockA{\IEEEauthorrefmark{4}School of Electrical and Electronic Engineering, University College Dublin, Ireland}
\IEEEauthorblockA{Email: moulichakraborty@ieee.org, avishek.nag@ucd.ie, SCHANDRA@tcd.ie,
anshu.mukherjee@ieee.org, \\\{k.koufos, m.dianati\}@qub.ac.uk}\vspace{-0.70cm}
\thanks{The authors acknowledge support from the Engineering and Physical Sciences Research Council (EPSRC) through the Quantum Position, Navigation and Timing (QEPNT) Hub (EP/Z533178/1).}}
\maketitle

\begin{abstract}
Satellite-based continuous-variable quantum key distribution (CV-QKD) is fundamentally constrained by finite communication windows, channel attenuation, and statistical fluctuations arising from a limited number of exchanged quantum states. This work develops a finite-size CV-QKD framework for satellite quantum communication under a hybrid noise model that jointly incorporates satellite geometry, atmospheric and propagation losses, quantum Poissonian noise coexisting with classical additive white Gaussian noise, detector imperfections, and composable finite-key security under collective attacks. A single canonical finite-size secret key rate (SKR) expression is derived in which the finite-size correction is stated in a strictly decreasing form in the block size, removing the ambiguity between the segmented and the block-level formulations used in earlier treatments. The finite-size accounting is further extended to the quantum bit error ratio (QBER) through an explicit parameter estimation penalty, so that the feasibility constraint inherits the same block-size dependence as the key rate. The framework quantifies the influence of reconciliation efficiency, transmission efficiency, detection efficiency, receiver aperture, beam waist, zenith angle, wavelength, and effective communication time. Numerical results reveal a significant performance gap between finite-size and asymptotic operation, indicating that asymptotic analysis can overestimate the feasible satellite altitude by a factor of several, and that the achievable range is governed by the number of signals actually exchanged rather than by any single physical channel parameter. These results establish block size as a first-class design variable for satellite CV-QKD mission dimensioning.
\end{abstract}


\section{Introduction}

Quantum key distribution (QKD) exploits superposition, the no cloning theorem, and measurement induced disturbance to let two parties establish a shared secret key with security guaranteed against an adversary of unbounded computational power, on the basis of physical law rather than computational hardness. Satellite QKD is a promising deployment path, since free space optical (FSO) links between satellites and ground stations overcome the range limitation of fibre-based terrestrial QKD and enable continental and global key distribution as a backbone for a future quantum internet \cite{ref1,ref20}. A central distinction in QKD performance analysis lies between the asymptotic and the finite size regimes. The asymptotic regime assumes an unbounded number of exchanged quantum states, so that statistical fluctuations vanish and idealised upper bound key rates are obtained; an assumption incompatible with satellite links where state exchange is limited by short pass durations. Finite size analysis instead accounts for transmission losses, hybrid channel noise, and the security parameter trade offs required by composable security proofs \cite{ref2,ref4}, since with only a finite number of symbols available for parameter estimation and key generation, the resulting statistical fluctuations render asymptotic predictions overly optimistic in practice \cite{ref3,ref25}.

Recent QKD research has moved from isolated point to point demonstrations towards integrated multi-scenario quantum secure infrastructure, spanning coexistence with passive optical networks \cite{ref5,ref6}, coherent state QKD with displacement receivers \cite{ref7,ref8}, unmanned aerial vehicle-assisted links \cite{ref9}, satellite to ground CV-QKD under turbulence \cite{ref20}, probabilistic amplitude shaping \cite{ref10}, trusted reconfigurable intelligent surfaces \cite{ref11}, group authentication \cite{ref12}, and multiple input multiple output terahertz channels \cite{ref13}. As QKD moves towards heterogeneous and mobility aware deployments, the channel and security models must evolve accordingly. This matters particularly for satellite quantum communication, where turbulence, background radiance, beam divergence, and fading produce dynamic and weather dependent fluctuations \cite{ref14,ref15}. Rather than idealising these impairments, finite size QKD folds their statistical variability into the SKR calculation, which is decisive for hybrid noise in which quantum Poissonian noise \cite{ref16} coexists with classical additive white Gaussian noise (AWGN) \cite{ref18} of thermal and electronic origin, an unavoidable combination in realistic receivers \cite{ref1}. Asymptotic models simplify this composition through a single source noise assumption, whereas the finite size treatment retains it in full \cite{ref19,ref20} and is further required for composable security \cite{ref21} and for extension to collective and coherent attacks \cite{ref23,ref24}.

The present treatment is delimited explicitly with respect to the authors' earlier work. The hybrid Poisson Gaussian noise decomposition and the associated Gaussian channel capacity were established in \cite{ref1} in the asymptotic regime and for a generic Gaussian channel, without satellite geometry. The satellite FSO channel model and its parametric characterisation were reported in \cite{ref3}, again under asymptotic key rate assumptions in which statistical fluctuations are absent by construction. Neither work carries a composable finite-key budget, neither treats block size as a design variable, and neither quantifies the discrepancy between the two regimes. The contribution of this paper is therefore the finite regime treatment itself, a single canonical finite-key expression for the hybrid noise satellite channel, a finite-size extension of the error rate constraint, and a direct quantification of the altitude range over which asymptotic dimensioning is misleading.

The main contributions of this paper are as follows,
\begin{itemize}
\item A hybrid quantum channel model integrating the fading characteristics of FSO satellite links, namely zenith angle, altitude, and effective communication time, with a hybrid Poisson Gaussian noise decomposition under collective attack security, from which a single canonical finite key expression is obtained with the correction term stated in a strictly decreasing form in the block size, removing the ambiguity between the segmented and the block level formulations used in earlier treatments.
\item A finite size extension of the error rate constraint, in which the parameter estimation penalty incurred by a finite sample is propagated through the effective signal to noise ratio (SNR), so that the quantum bit error ratio (QBER) inherits the block size dependence of the key rate and the feasibility test is stated consistently with the composable security budget.
\item Quantification of the influence of reconciliation efficiency, transmission coefficient, detection efficiency, telescope aperture, beam waist, zenith angle, and communication time, together with the trade offs induced by altitude and wavelength, evaluated against a direct finite size versus asymptotic comparison that quantifies the practical performance gap between the two regimes.
\end{itemize}

\textit{Notation:} $\mathrm{tr}$ denotes the trace of a matrix and $T(\cdot)$ a trace preserving map. The adjoint of $A$ is $A^{\dagger}$, $\otimes$ is the tensor product, $\lambda$ the Poisson parameter, and $\mathcal{N}$ the Gaussian density. Throughout, $\tau$ denotes the instantaneous transmission efficiency, $\theta$ the zenith angle, and $\theta_{\mathrm{pt}}$ the combined pointing error angle. To avoid symbol collision, $\xi$ denotes the channel excess noise and $\epsilon$ with a subscript is reserved exclusively for composable security parameters; the symplectic eigenvalues are written $\nu_i$ and the wavelength $\omega$, so that $\lambda$ and $w_0$ retain their single meanings as the Poisson parameter and the beam waist respectively. Turbulence strength is denoted $\sigma_{\tau}$ and the modulation standard deviation $\sigma_X$.

\section{Satellite Channel and Hybrid Noise Model}

\begin{figure*}[t]
\centering
\includegraphics[width=0.32\textwidth]{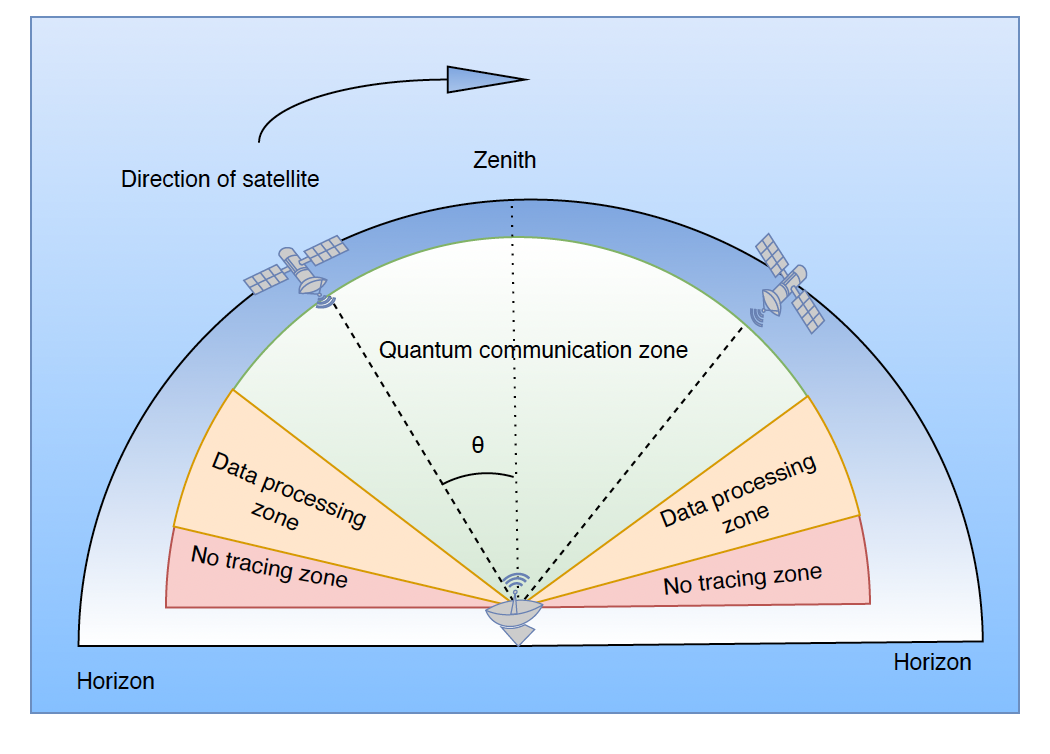}\hfill
\includegraphics[width=0.32\textwidth]{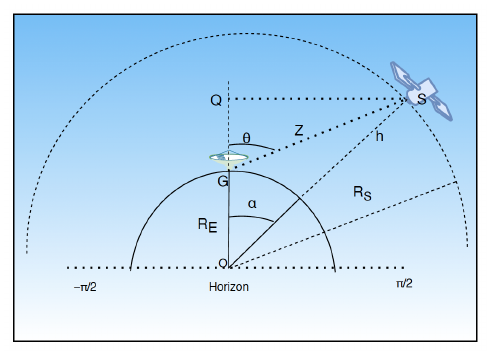}\hfill
\includegraphics[width=0.3\textwidth]{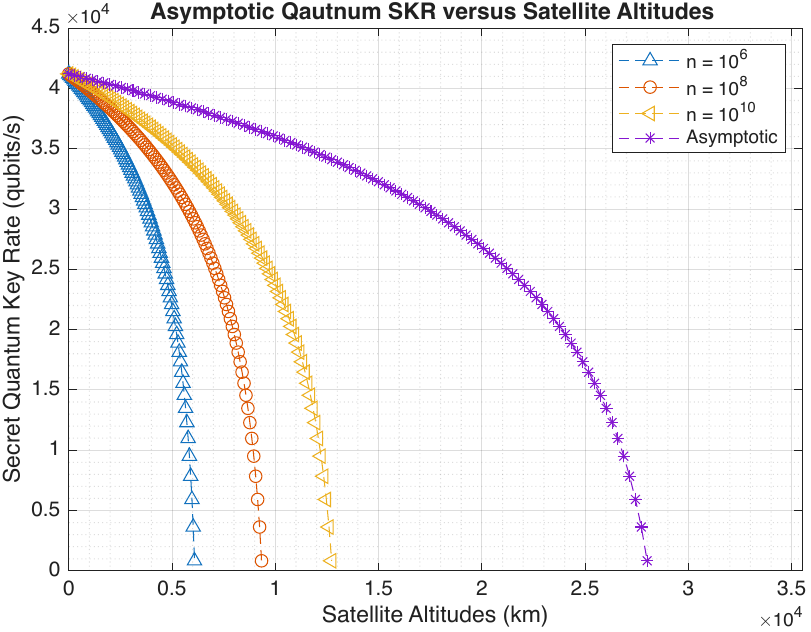}
\caption{(a) Geometric configuration of the satellite to ground link, showing the parameters that govern slant range, visibility window, and channel transmissivity, (b) Operational zones of a satellite pass for quantum communication, showing the quantum communication zone around zenith, adjacent data processing zones, and no tracing regions near the horizon, (c) Asymptotic SKR versus finite size SKR against satellite altitude for block sizes $m = 10^6, 10^8, 10^{10}$, and the asymptotic case at $M = 10^{20}$.}
\label{fig1}\vspace{-0.25cm}
\end{figure*}

Fig.~\ref{fig1}(a) depicts the link geometry, with a ground station near sea level ($h_0 \approx 0$) and a satellite at altitude $h > 100$ km, the zenith angle $\theta$ ranging from $0$ at overhead to $\pi/2$ at the horizon. Orbital geometry determines the slant distance $Z$ as a function of altitude $h$, zenith angle $\theta$, orbital radius $R_S$, and Earth radius $R_E$, and therefore governs channel transmissivity,
\begin{equation}
Z(h,\theta) = \sqrt{R_S^2 + R_G^2\left(\cos^2\theta - 1\right)} - R_G\cos\theta,
\label{eq1}
\end{equation}
where $R_G = R_E + h_0$ and $R_S = R_E + h$, with $h_0 = 0$ and $R_E = 6371$ km. For circular orbits at zenith crossing this reduces to $Z(R_S,\alpha) = \sqrt{R_G^2 + R_S^2 - 2R_GR_S\cos\alpha}$, where $\alpha = 2\pi t/T_S$ is the orbital angle for elapsed time $t$ and $T_S = 2\pi\sqrt{R_S^3/GM_E}$ is the orbital period. The time required to reach zenith angle $\theta$ at altitude $h$ is
\begin{equation}
t(\theta,h) = \sqrt{\frac{(R_E+h)^3}{GM_E}}\arccos\left(\frac{R_G + Z(h,\theta)\cos\theta}{R_E+h}\right),
\label{eq2}
\end{equation}
which yields the total effective communication time \cite{ref22}
\begin{equation}
t_{\mathrm{total}} \approx \frac{2(R_E+h)^{3/2}}{\sqrt{GM_E}}\left[\theta_{\max} - \arcsin\left(\frac{R_E\sin\theta_{\max}}{R_E+h}\right)\right],
\label{eq3}
\end{equation}
with $\theta_{\max}$ the maximum admissible zenith angle. As illustrated in Fig.~\ref{fig1}(b), the zenith angle divides the pass into a good sector ($-1 \leq \theta \leq 1$) of optimal transmissivity, a data processing sector ($1 < \theta < \pi/2$), and an encrypted communication sector ($-\pi/2 < \theta < -1$) relying on previously established keys, so that quantum state transmission is performed around zenith where the propagation path is shortest and the subsequent processing is carried out towards the edges of the window.

The quantum channel is modelled as a completely positive trace preserving map $\rho \mapsto T(\rho)$ on density operators in a Hilbert space $\mathcal{H}$ \cite{ref26}, for which there exist an environment state $\rho_E$ and a unitary $U$ with $T(\rho) = \mathrm{tr}_E[U(\rho \otimes \rho_E)U^{\dagger}]$. Alice draws $2M$ real Gaussian variables of variance $\sigma_X^2$ and maps them onto $M$ coherent states through $\alpha_m = X_{2m-1} + iX_{2m}$, while Bob performs homodyne detection under reverse reconciliation (RR), giving the $m$th use of the FSO satellite channel as \cite{ref27},
\begin{equation}
Y_m = \sqrt{\tau_m}\,X_m + N_m,
\label{eq4}
\end{equation}
where $N_m = N_m^{(1)} + N_m^{(2)}$ combines quantum Poissonian noise $N_m^{(1)}$ of probability mass function $f_{N_m^{(1)}}(j) = e^{-\lambda}\lambda^j/j!$ with classical AWGN $N_m^{(2)}$ of mean $\mu_{N_m^{(2)}}$ and variance $\sigma_{N_m^{(2)}}^2$. With the Poisson weight $p_j = e^{-\lambda}\lambda^j/j!$, the resulting hybrid density is \cite{ref1}
\begin{equation}
f_{N_m}(z) = \sum_{j=0}^{\infty} p_j \frac{1}{\sigma_{N_m^{(2)}}\sqrt{2\pi}}\,
e^{-\frac{1}{2}\left(\frac{z-j-\mu_{N_m^{(2)}}}{\sigma_{N_m^{(2)}}}\right)^2}.
\label{eq5}
\end{equation}
The received signal is thus the convolution of the Gaussian modulated transmitted signal with the hybrid noise, and the per use channel capacity can be approximated as \cite{ref3}
\begin{align}
C_m = \sum_{j=0}^{R} p_j \Bigg[ &-\log_2 p_j + \frac{1}{2}\log_2\!\left(2\pi e\,\tau\sigma_{X_m}^2 + \sigma_{N_m^{(2)}}^2\right) \nonumber\\
&+ \log_2 \sum_{l=0}^{R} p_l \frac{e^{-(j-l)^2/4\sigma_{N_m^{(2)}}^2}}{2\sigma_{N_m^{(2)}}\sqrt{\pi}} \Bigg],
\label{eq6}
\end{align}
where $R$ is the truncation order of the Poisson series.

The transmission efficiency $\tau \in [0,1]$ is a random variable whose distribution is well approximated by a lognormal law \cite{ref3},
\begin{equation}
P(\tau_m) = \frac{1}{\sqrt{2\pi}\,\sigma_{\tau}\,\tau_m}\,
e^{-\frac{1}{2}\left(\frac{\ln \tau_m + \ln \tilde{\tau}_m}{\sigma_{\tau}}\right)^2},
\label{eq7}
\end{equation}
where $-\ln\tilde{\tau}_m$ is the log mean transmittance and $\sigma_{\tau}$ characterises turbulence. This model captures the weak to moderate turbulence regime typical of near vertical links while remaining analytically tractable, whereas Gamma Gamma models suit strong turbulence over long horizontal paths. Beam divergence sets the spot size $w_g(Z) = w_0\sqrt{1 + (Z/N_R)^2}$ with Rayleigh range $N_R = \pi w_0^2/\omega$, beam waist $w_0$, and wavelength $\omega$, so the aperture collection efficiency for receiver radius $a_r$ is $\eta_g(Z) = 1 - e^{-2a_r^2/w_g^2}$. Atmospheric extinction obeys the Beer Lambert law $\eta_{\mathrm{atm}}(h) = e^{-\delta(h)Z}$ with $\delta(h) = \delta_0 e^{-h/\tilde{h}}$, $\tilde{h} = 6600$ m and $\delta_0 = 5\times10^{-6}$ m$^{-1}$ at $\omega = 800$ nm, giving zenith transmittance $\eta_{\mathrm{zen}} = 0.94$ and elevation scaling $\eta_{\mathrm{atm}} = \eta_{\mathrm{zen}}^{\sec(\theta)}$. Pointing and tracking losses are captured by $\eta_{\mathrm{pt}} = \exp(-G_{tx}\theta_{\mathrm{pt}}^2)$ with transmit gain $G_{tx} = (2\pi a_S/\omega)^2$, and detector gating over a window $\Delta t$ contributes $\eta_{\mathrm{temporal}} = 1 - e^{-\Delta t/t}$, so that the total transmission efficiency is
\begin{equation}
\tau = \eta_{\mathrm{det}}\,\eta_g\,\eta_{\mathrm{atm}}\,\eta_{\mathrm{pt}}\,\eta_{\mathrm{temporal}},
\label{eq8}
\end{equation}
which enters the key rate multiplicatively and dominates the altitude dependence reported in Section~\ref{sec:results}. 


\vspace{-0.2cm}
\section{Secret Key Rate in the Finite Regime}
\label{sec:skr}
\subsection{Asymptotic Reference}

Under RR and collective attacks, the asymptotic SKR follows the Devetak Winter bound \cite{ref20,ref3}
\begin{multline}
    K = \lim_{M\to\infty}\left\{\frac{1}{M}\left[H(Y^{(M)}) - \mathrm{leak}_{\mathrm{EC}}^{(M)}\right] - \Psi(\tilde{\Gamma}^{(M)})\right\} \\ = \beta I_{AB} - \chi_{BE},
\label{eq9}
\end{multline}
where $H(Y^{(M)})$ is the empirical entropy of Bob's data, $\mathrm{leak}_{\mathrm{EC}}^{(M)}$ is the information disclosed during error correction, and $\Psi(\tilde{\Gamma}^{(M)})$ bounds Eve's accessible information. With Gaussian modulation and reconciliation efficiency $\beta$, the mutual information term is $\beta I_{AB} = \beta M^{-1}\sum_{m=1}^{M} C_m$. Beyond the hybrid noise, the channel carries excess noise $\xi$ and detector imperfections, so that $\chi_{\mathrm{line}} = 1/\sqrt{\tau} - 1 + \xi$, $\chi_{\mathrm{hom}} = (1 + \vartheta_{\mathrm{ele}})/\eta_{\mathrm{det}} - 1$, and $\chi_{\mathrm{tot}} = \chi_{\mathrm{line}} + \chi_{\mathrm{hom}}/\sqrt{\tau}$. Eve's information is the Holevo bound
\begin{equation}
\chi_{BE} = \sum_{i=1}^{2} G\left(\frac{\nu_i - 1}{2}\right) - \sum_{i=3}^{4} G\left(\frac{\nu_i - 1}{2}\right),
\label{eq10}
\end{equation}
with $G(x) = (x+1)\log_2(x+1) - x\log_2 x$ and $\nu_i$ the symplectic eigenvalues of the covariance matrix of the state shared between Bob and a purification of Eve's system.

\subsection{Canonical Finite Key Expression}

For a block of $M$ pulses, $M - n$ signals are consumed by parameter estimation and $n$ are retained for key generation. The finite key rate under RR and collective attacks is stated here in the single canonical form
\begin{equation}
K_{\mathrm{fin}}(\tau,n) = \beta I_{AB}(\tau) - \chi_{BE}(\tau) - \Delta(n),
\label{eq11}
\end{equation}
subject to $\mathrm{QBER}_{\mathrm{fin}}(\tau,n) \leq \mathrm{QBER}_{\max}$, where $\Delta(n)$ is the finite size correction associated with the smooth min entropy and with the failure probability of privacy amplification,
\begin{equation}
\Delta(n) = \left(2\dim\mathcal{H}_X + 3\right)\sqrt{\frac{\log_2(2/\bar{\epsilon})}{n}} + \frac{2}{n}\log_2\left(\frac{1}{\epsilon_{PA}}\right).
\label{eq12}
\end{equation}
Here $\dim\mathcal{H}_X = 2$ for the raw key quadrature variable. Both terms of \eqref{eq12} decrease monotonically in $n$, so $\Delta(n)\to 0$ and $K_{\mathrm{fin}} \to K$ as $n \to \infty$, which is the property required for a well posed comparison between the two regimes. The compact approximation $\Delta(n) \approx 7\sqrt{\log_2(2/\bar{\epsilon})/n}$ is recovered from \eqref{eq12} by neglecting the privacy amplification term, and is stated here only to identify the leading $n^{-1/2}$ scaling; all reported results use the full form \eqref{eq12}.

The protocol is $\epsilon_{\mathrm{tot}}$ secure in the composable sense with
\begin{equation}
\epsilon_{\mathrm{tot}} = \epsilon_{\mathrm{EC}} + \epsilon_{PE} + \epsilon_{PA} + \bar{\epsilon},
\label{eq13}
\end{equation}
where $\epsilon_{\mathrm{EC}}$ is the error correction failure probability, $\epsilon_{PE}$ the parameter estimation failure probability, $\epsilon_{PA}$ the privacy amplification failure probability, and $\bar{\epsilon}$ the smoothing parameter. Each term is set to $10^{-10}$ in the evaluation, giving $\epsilon_{\mathrm{tot}} = 4\times10^{-10}$, which balances security against practical feasibility. Only $\bar{\epsilon}$ and $\epsilon_{PA}$ enter \eqref{eq12} explicitly; $\epsilon_{PE}$ enters through the error rate constraint of Section~\ref{sec:qber} and $\epsilon_{\mathrm{EC}}$ through $\mathrm{leak}_{\mathrm{EC}}$, so that the budget is complete and no failure probability is counted twice.

When a pass is partitioned into $S$ quasi static segments of raw key length $n_i$, the aggregate rate follows from \eqref{eq11} as a weighted sum in which only positive contributions are retained \cite{ref17},
\begin{equation}
K_{\mathrm{fin}}^{\mathrm{pass}} = \sum_{i=1}^{S} \max\left\{0, \frac{n_i}{M}\left[\beta I_{AB} - \chi_{BE} - \Delta(n_i)\right]\right\},
\label{eq14}
\end{equation}
with $\sum_i M_i = M$, so segments whose transmissivity is too low do not penalise the aggregate rate. Expression \eqref{eq14} is the segmented generalisation of \eqref{eq11}.  For every bipartite state $\rho_{AB}$ it holds that $K_{\mathrm{fin}}(\rho_{AB}) \geq K_{\mathrm{fin}}(\rho_{AB}^{G})$, so the Gaussian evaluation is a valid lower bound under collective attacks, and it is tight in the Gaussian CV setting where Eve's optimal attack is itself Gaussian.

\subsection{Quantum Bit Error Rate in the Finite Regime}
\label{sec:qber}

For Gaussian modulated coherent state CV-QKD with homodyne detection, the error probability of the sifted key decision variable is governed by the effective SNR at Bob's detector, $\mathrm{SNR}_{\mathrm{eff}}(\tau) = \tau V_A/(1 + \chi_{\mathrm{tot}}(\tau))$ with $V_A$ the modulation variance, and follows the Gaussian channel error function form
\begin{equation}
\mathrm{QBER}(\tau) = \tfrac{1}{2}\,\mathrm{erfc}\!\left(\sqrt{\mathrm{SNR}_{\mathrm{eff}}(\tau)/2}\right).
\label{eq15}
\end{equation}
Expression \eqref{eq15} carries no block size dependence and therefore describes the asymptotic error rate only. In the finite regime the transmittance is not known but estimated from the $(M - n)$ disclosed samples, and the worst case value consistent with the confidence interval must be used. Writing the estimator dispersion as $\sigma_{\hat{\tau}} = c_{\tau}\tau/\sqrt{n}$ with $c_{\tau}$ a channel dependent constant, and $z_{\epsilon_{PE}}$ for the standard normal quantile at confidence $1 - \epsilon_{PE}$, the admissible lower transmittance is $\tau_{\mathrm{low}} = \tau(1 - c_{\tau}z_{\epsilon_{PE}}/\sqrt{n})$. Propagating $\tau_{\mathrm{low}}$ through \eqref{eq15} and expanding to first order in $n^{-1/2}$ gives the finite size error rate
\begin{equation}
\mathrm{QBER}_{\mathrm{fin}}(\tau,n) = \mathrm{QBER}(\tau) + \frac{\kappa(\tau)}{\sqrt{n}}, \quad
\label{eq16}
\end{equation}
with $\kappa(\tau) = -c_{\tau}z_{\epsilon_{PE}}\,\tau\,\frac{\partial\,\mathrm{QBER}}{\partial\tau},$ $\kappa(\tau) \geq 0$ since $\mathrm{QBER}$ is monotonically decreasing in $\tau$. Expression \eqref{eq16} is the constraint applied in \eqref{eq11}, it recovers \eqref{eq15} as $n \to \infty$, and it shares the $n^{-1/2}$ scaling of the leading term of \eqref{eq12}, so that key rate and error rate degrade under the same statistical mechanism. The QBER also enters the reconciliation budget and the mutual information term implicitly through $\chi_{\mathrm{tot}}$, and is reported alongside $K_{\mathrm{fin}}$ as the feasibility constraint.

\begin{figure*}[t]
\centering
\includegraphics[width=0.33\textwidth]{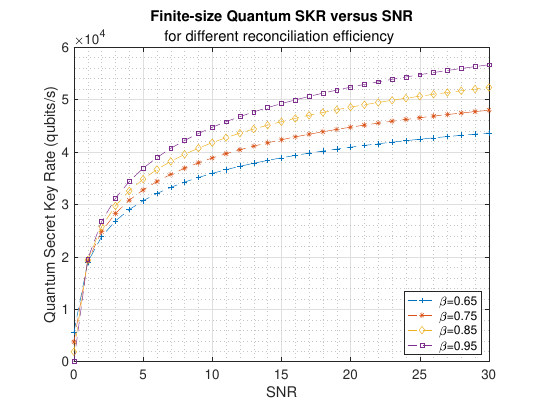}\hfill
\includegraphics[width=0.28\textwidth]{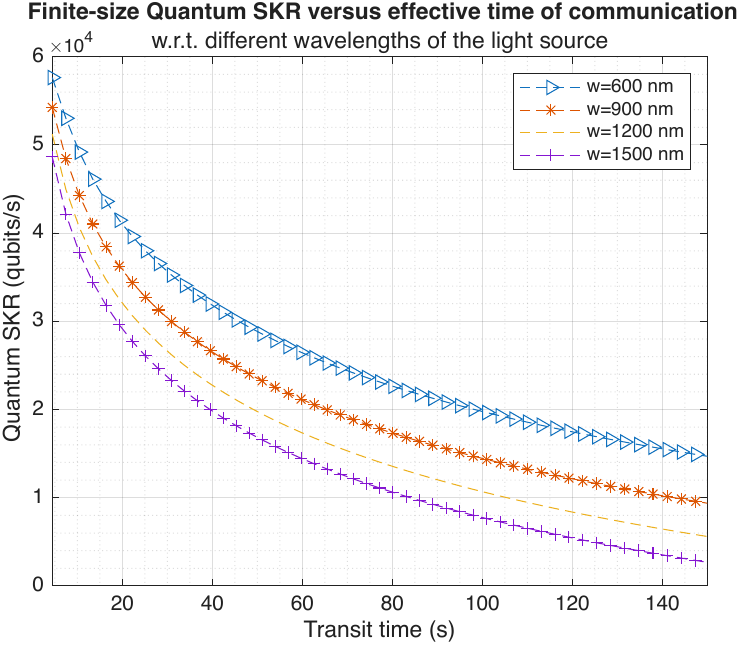}\hfill
\includegraphics[width=0.33\textwidth]{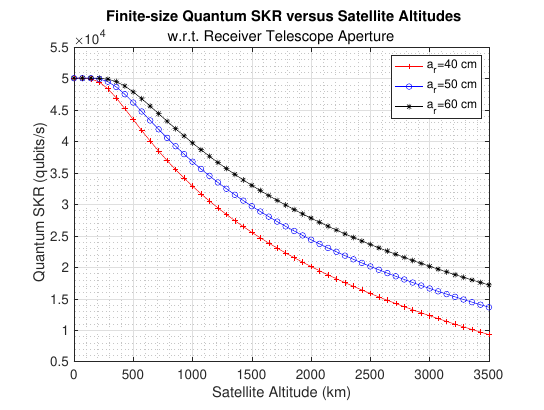}
\caption{(a) Finite size SKR versus SNR for reconciliation efficiency $\beta$ from 0.65 to 0.95, (b) Dependence of the finite size SKR on the effective communication time for light source wavelengths from 600 nm to 1500 nm, (c) Finite size SKR versus satellite altitude for receiver telescope apertures $a_r$ from 40 cm to 60 cm.}
\label{fig2}\vspace{-0.25cm}
\end{figure*}

\section{Numerical Results and Discussion}
\label{sec:results}
Unless stated otherwise, the evaluation uses the parameters with excess noise $\xi = 0.35$ SNU for the altitude, beam waist, zenith angle, and detection efficiency sweeps, and $\xi = 0.63$ SNU for the reconciliation efficiency, transit time, and aperture sweeps. All key rates are expressed in bits per second and are obtained from the secure fraction per channel use multiplied by the symbol rate $R_s$, so that the reported ordinate scales linearly with $R_s$ and the relative comparisons between regimes are independent of it. The transmission efficiency decreases from 1 to 0.3 with increasing altitude in accordance with the fading model of Section II.

Fig.~\ref{fig1}(c) reports the central comparison of this work. Both regimes exhibit the same qualitative decay with altitude, since rising altitude reduces $\tau$ through atmospheric attenuation, beam divergence, and propagation loss, and therefore limits the number of states reaching the receiver. The two regimes differ sharply in the altitude at which the rate collapses. At $m = 10^6$, a block size representative of a single short pass, the key rate vanishes near $7\times10^3$ km, whereas the asymptotic curve remains positive beyond $3\times10^4$ km. Increasing the block size to $10^8$ and $10^{10}$ recovers part of this margin, consistent with the $n^{-1/2}$ scaling of the leading term of \eqref{eq12}. The practical implication is direct, a mission dimensioned from asymptotic curves would overestimate the usable altitude range by several kilometres, and the discrepancy is governed by the number of signals actually exchanged rather than by any physical channel parameter.

Fig.~\ref{fig2} illustrates the sensitivity of the finite size SKR to three design parameters. In Fig.~\ref{fig2}(a), a higher reconciliation efficiency reduces the information discarded during error correction and raises the SKR across the whole SNR range. The gain between $\beta = 0.65$ and $\beta = 0.95$ exceeds $10^4$ bits/s at high SNR; since $\beta$ scales the mutual information term in \eqref{eq11} while leaving $\chi_{BE}$ unchanged, reconciliation performance maps almost linearly onto the key rate and is the most cost-effective in network level. Fig.~\ref{fig2}(b) relates the SKR to the effective communication time and wavelength, where the zenith angle and altitude restrict the usable window to less than the total transit time. Longer wavelengths admit more noise photons and degrade state fidelity, contracting the effective time window until the 1500 nm curve crosses zero before 150 s. Shorter wavelengths suppress noise but incur higher atmospheric attenuation, so the wavelength must be selected jointly with the link budget. Fig.~\ref{fig2}(c) shows that a larger receiver aperture, with $a_r$ between 40 cm and 60 cm, improves photon collection and reduces loss. The benefits from larger apertures grow with altitude as the slant range and received spot size increase; at 3500 km the rate for $a_r = 60$ cm is approximately twice that for $a_r = 40$ cm, identifying the aperture as the dominant hardware level parameter at high altitude, subject to the pointing stability a larger telescope demands.

\begin{figure*}[t]
\centering
\includegraphics[width=0.28\textwidth]{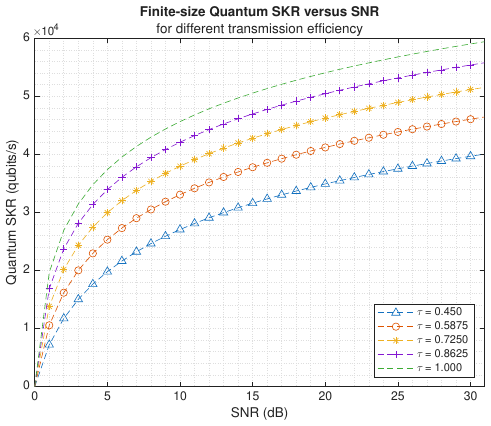}\hfill
\includegraphics[width=0.33\textwidth]{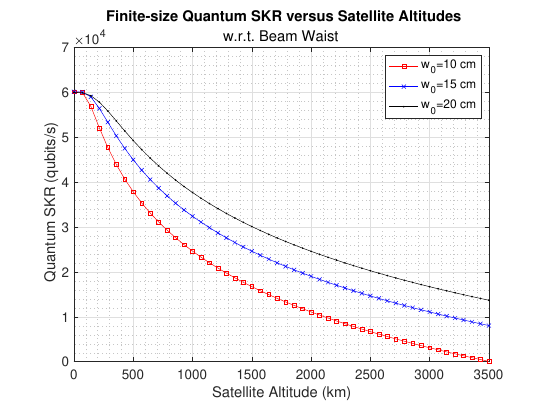}\hfill
\includegraphics[width=0.33\textwidth]{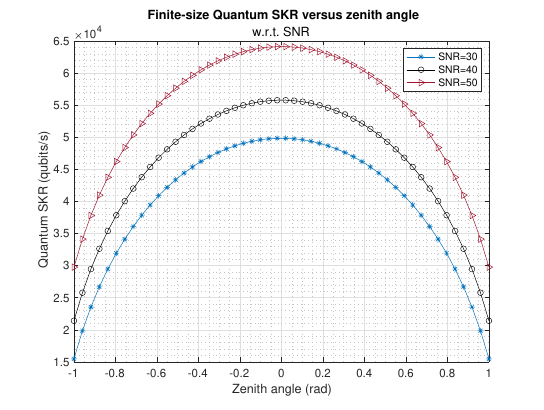}
\caption{(a) Finite size SKR versus SNR for transmission efficiency $\tau$ between 0.450 and 1.000, (b) Finite size SKR versus satellite altitude for beam waist $w_0$ between 10 cm and 20 cm, (c) Finite regime SKR versus zenith angle within the good sector for SNR values of 30, 40, and 50 dB.}
\label{fig3}\vspace{-0.25cm}
\end{figure*}

Fig.~\ref{fig3} isolates three further channel level parameters. In Fig.~\ref{fig3}(a) the transmission efficiency enters the key rate multiplicatively, and reducing $\tau$ from 1.000 to 0.450 costs approximately $2\times10^4$ bits/s at 30 dB SNR, so a modest error in the estimated transmittance propagates directly into the predicted rate, which is precisely the mechanism formalised in \eqref{eq16}. Fig.~\ref{fig3}(b) sweeps the beam waist for $w_0$ between 10 cm and 20 cm.  All configurations coincide at low altitude, where the aperture captures essentially the full beam, and separate progressively as altitude grows, since a larger waist mitigates diffraction spreading at long slant range while excessively large waists introduce aberration and alignment penalties, so an interior optimum exists for any given aperture and altitude profile. Fig.~\ref{fig3}(c) presents the zenith angle dependence within the good sector, $-1$ to $1$ rad, where the rate peaks at zenith with the shortest slant range and minimal attenuation and decays symmetrically towards the sector edges as longer paths, diffraction, scattering, and absorption reduce the delivered signal. Raising the SNR from 30 to 50 lifts the entire profile by approximately $1.5\times10^4$ bits/s without altering its shape, indicating that link budget improvements and geometric scheduling act as independent levers.

\begin{figure*}[t]
\centering
\includegraphics[width=0.30\textwidth]{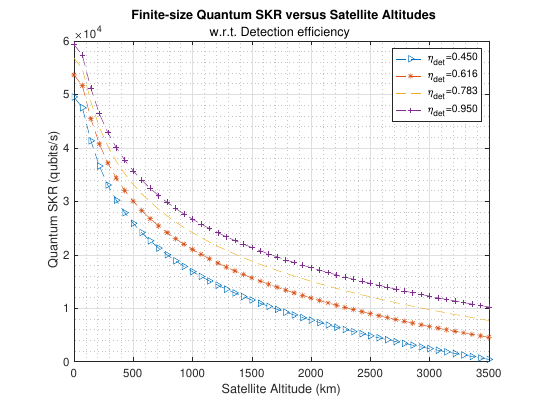}\hfill
\includegraphics[width=0.26\textwidth]{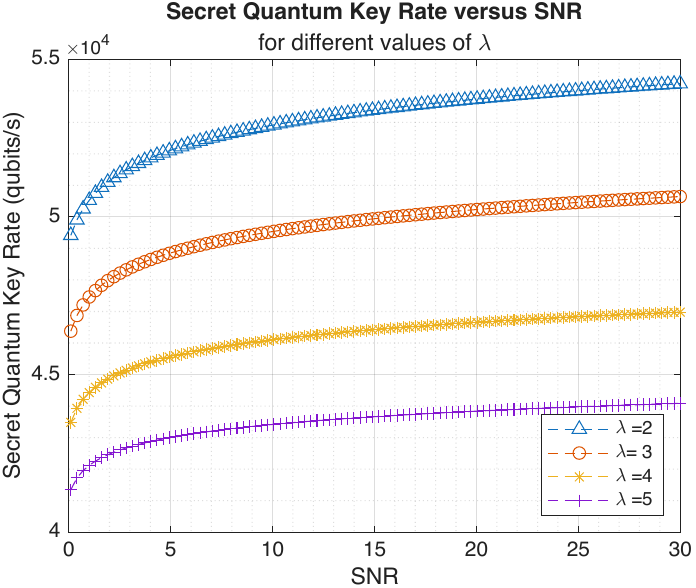}
\hfill
\includegraphics[width=0.30\textwidth]{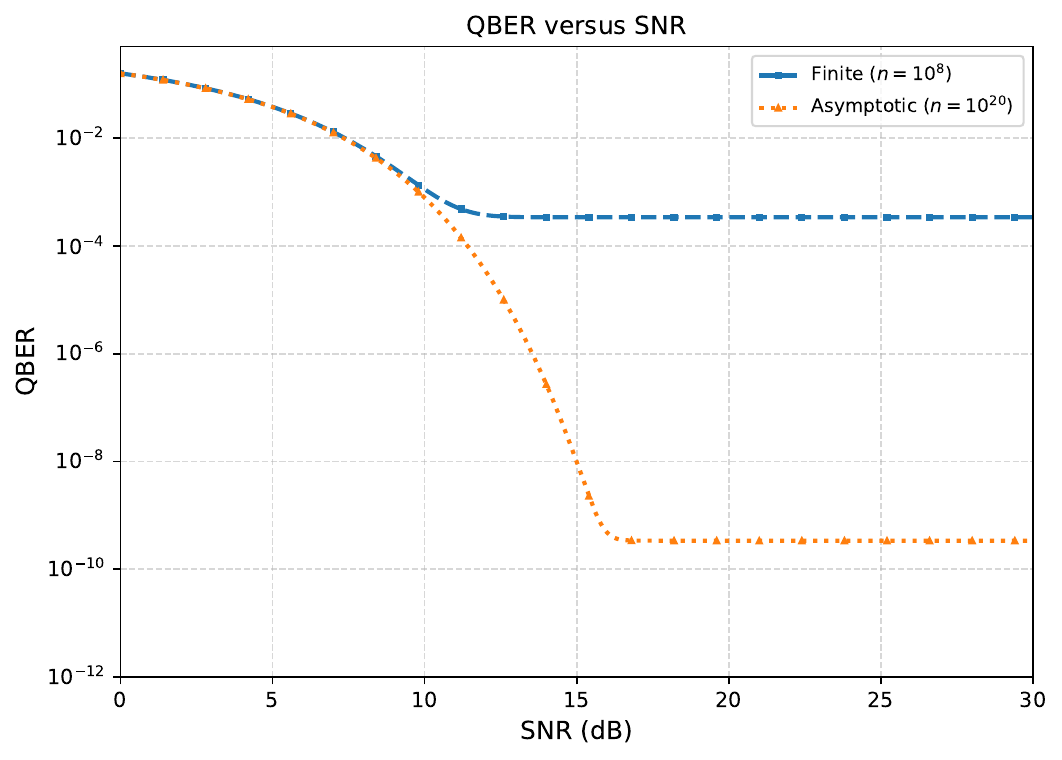}
\caption{(a) Impact of detection efficiency $\eta_{\mathrm{det}}$ on the finite regime SKR at different satellite altitudes, (b) SKR versus SNR for different Poisson noise parameters $\lambda=2,3,4,$ and $5$ under the hybrid noise model, (c) QBER versus SNR for the finite size case ($n = 10^{8}$) and the asymptotic case ($M = 10^{20}$).}
\label{fig4}\vspace{-0.25cm}
\end{figure*}

Fig.~\ref{fig4}(a) completes the parametric study with the detection efficiency sweep for $\eta_{\mathrm{det}}$ between 0.450 and 0.950, whose influence persists across the whole altitude range rather than concentrating at either end; taken together with Figs.~\ref{fig1}(c) to \ref{fig4}(a), this identifies $\beta$, $\tau$, and $\eta_{\mathrm{det}}$ as the parameters requiring the tightest control, since each acts on a distinct mechanism, namely information extraction, photon delivery, and photon registration, and no single one compensates for a deficit in another. Fig.~\ref{fig4}(b) shows that the SKR increases with SNR for all considered $\lambda$, while increasing the Poisson parameter reduces the achievable SKR due to the higher quantum noise contribution.  Fig.~\ref{fig4}(c) illustrates the QBER versus SNR, where the finite and asymptotic curves remain close at low SNR because performance is limited predominantly by physical channel errors. The QBER analysis confirms that channel degradation and finite-size effects directly limit the reliability of satellite CV-QKD, with higher noise levels and increasing satellite altitude leading to larger error rates. At 20 dB, the finite size case with $m = 10^{8}$ achieves a QBER of approximately $3.39\times10^{-4}$, against asymptotic values on the order of $10^{-8}$. The four order of magnitude separation confirms that at high SNR the residual error rate is set by the finite sample rather than by the channel, and therefore that large block operation is the only effective remedy once the link budget has been optimised. Taken together with Fig.~\ref{fig1}(c), this establishes that the block size governs both the feasibility constraint and the achievable rate through the same $n^{-1/2}$ statistical mechanism, and is therefore not a post hoc correction to an asymptotic design but a primary dimensioning variable.
\section{Conclusion}

This work developed a finite size framework for satellite based CV-QKD under a hybrid Poisson Gaussian noise model, incorporating finite key effects together with satellite geometry, channel attenuation, detector imperfections, reconciliation efficiency, receiver aperture, beam waist, zenith angle, wavelength, and effective communication time. A single canonical finite key expression was obtained with a strictly decreasing correction term. The composable security budget was stated in full, and the finite size accounting was extended to the error rate constraint through an explicit parameter estimation penalty sharing the same $n^{-1/2}$ scaling. The numerical results show that asymptotic analysis can significantly overestimate the achievable operating range, since finite size penalties become dominant when the number of exchanged quantum states is limited by the satellite visibility window, so that the usable altitude depends strongly on the available block size. The parametric study further identifies reconciliation efficiency, transmission efficiency, and detection efficiency as the parameters requiring the tightest control, each acting through a distinct and non substitutable mechanism. These results demonstrate that finite size security analysis must be embedded at the design stage of satellite CV-QKD missions rather than applied as a correction to asymptotic predictions, and provide a foundation for future work on adaptive block size selection along a pass, empirical validation against deployed links, and multi node scheduling for scalable quantum infrastructure.
\balance


\begin{thebibliography}{99}
\bibitem{ref1} M. Chakraborty \textit{et al.}, ``Hybrid quantum noise model to compute Gaussian quantum channel capacity,'' \textit{IEEE Access}, vol. 12, pp. 14\,671--14\,689, 2024.
\bibitem{ref2} X. Zhang \textit{et al.}, ``Finite-size analysis of continuous-variable measurement-device-independent quantum key distribution,'' \textit{Phys. Rev. A}, vol. 96, no. 4, p. 042334, 2017.
\bibitem{ref3} M. Chakraborty \textit{et al.}, ``A hybrid noise approach to modeling of free-space satellite quantum communication channel for continuous-variable QKD,'' \textit{IEEE Trans. Green Commun. Netw.}, vol. 9, no. 3, pp. 1311--1325, 2025.
\bibitem{ref4} J. S. Sidhu \textit{et al.}, ``Finite key effects in satellite quantum key distribution,'' \textit{npj Quantum Inf.}, vol. 8, no. 1, p. 18, 2022.
\bibitem{ref5} M. Chakraborty \textit{et al.}, ``Hybrid quantum noise approximation and pattern analysis on parameterized component distributions,'' in \textit{Proc. IEEE MECOM}, 2024, pp. 251--256.
\bibitem{ref6} A. Gagliano \textit{et al.}, ``Quantum key distribution spectral allocation and performance in coexistence with passive optical network standards,'' \textit{IEEE Trans. Commun.}, vol. 73, no. 1, pp. 510--523, 2025.
\bibitem{ref7} M. Chakraborty \textit{et al.}, ``An unsupervised machine learning to optimize hybrid quantum noise clusters for Gaussian quantum channel,'' \textit{IEEE Trans. Green Commun. Netw.}, 2025.
\bibitem{ref8} M. Zhao \textit{et al.}, ``Security of coherent-state quantum key distribution using displacement receiver,'' \textit{IEEE J. Sel. Areas Commun.}, vol. 42, no. 7, pp. 1871--1884, 2024.
\bibitem{ref9} P.-Y. Kong, ``UAV-assisted quantum key distribution for secure communications with resource limited devices,'' \textit{IEEE Trans. Veh. Technol.}, vol. 73, no. 8, pp. 11\,923--11\,933, 2024.
\bibitem{ref10} M. N. Notarnicola \textit{et al.}, ``Probabilistic amplitude shaping for continuous-variable quantum key distribution with discrete modulation over a wiretap channel,'' \textit{IEEE Trans. Commun.}, vol. 72, no. 1, pp. 375--386, 2024.
\bibitem{ref11} S. Kisseleff \textit{et al.}, ``Trusted reconfigurable intelligent surface for multiuser quantum key distribution,'' \textit{IEEE Commun. Lett.}, vol. 27, no. 8, pp. 2237--2241, 2023.
\bibitem{ref12} Y. Luo \textit{et al.}, ``An information-theoretic secure group authentication scheme for quantum key distribution networks,'' \textit{IEEE Trans. Commun.}, vol. 71, no. 9, pp. 5420--5431, 2023.
\bibitem{ref13} N. K. Kundu \textit{et al.}, ``Channel estimation and secret key rate analysis of MIMO terahertz quantum key distribution,'' \textit{IEEE Trans. Commun.}, vol. 70, no. 5, pp. 3350--3363, 2022.
\bibitem{ref14} M. Chakraborty, ``Hybrid quantum noise model for Gaussian quantum channels: Development, optimization, and applications,'' Ph.D. dissertation, Univ. of Dublin, 2025.
\bibitem{ref15} A. Schranz \textit{et al.}, ``A portable ambient optical noise measurement station,'' in \textit{Proc. ICTON}, 2023, pp. 1--6.
\bibitem{ref16} A. M. Fox, \textit{Quantum Optics: An Introduction}. Oxford Univ. Press, 2006.
\bibitem{ref17} M. Chakraborty, A. Mukherjee, T. Q. Duong, A. Nag, and S. Chandra, ``A hybrid noise model for satellite quantum communication channels: Continuous-variable quantum key distribution in the finite-size regime,'' \textit{TechRxiv}, 2025.
\bibitem{ref18} J. Oh \textit{et al.}, ``Continuous-variable quantum key distribution with time-division dual-quadrature homodyne detection,'' \textit{Opt. Express}, vol. 31, no. 19, pp. 30\,669--30\,681, 2023.
\bibitem{ref19} M. Chakraborty \textit{et al.}, ``Comparative analysis of differential and collision entropy for finite-regime QKD in hybrid quantum noisy channels,'' in \textit{Proc. IEEE WCNC}, 2026, pp. 1--6.
\bibitem{ref20} D. Dequal \textit{et al.}, ``Feasibility of satellite-to-ground continuous-variable quantum key distribution,'' \textit{npj Quantum Inf.}, vol. 7, no. 1, p. 3, 2021.
\bibitem{ref21} I. B. Djordjevic, \textit{Quantum Communication, Quantum Networks, and Quantum Sensing}. Academic Press, 2022.
\bibitem{ref22} N. Hosseinidehaj \textit{et al.}, ``Satellite-based continuous-variable quantum communications: State-of-the-art and a predictive outlook,'' \textit{IEEE Commun. Surveys Tuts.}, vol. 21, no. 1, pp. 881--919, 2019.
\bibitem{ref23} I. B. Djordjevic, ``Hybrid CV-DV quantum communications and quantum networks,'' \textit{IEEE Access}, vol. 10, pp. 23\,284--23\,292, 2022.
\bibitem{ref24} M. Curty \textit{et al.}, ``Finite-key analysis for measurement-device-independent quantum key distribution,'' \textit{Nat. Commun.}, vol. 5, no. 1, p. 3732, 2014.
\bibitem{ref25} T. Islam \textit{et al.}, ``Finite-resource performance of small-satellite-based quantum-key-distribution missions,'' \textit{PRX Quantum}, vol. 5, no. 3, p. 030101, 2024.
\bibitem{ref26} N. J. Cerf \textit{et al.}, \textit{Quantum Information with Continuous Variables of Atoms and Light}. Imperial College Press, 2007.
\bibitem{ref27} M. Chakraborty \textit{et al.}, ``On the Achievable Rate of Satellite Quantum Communication Channel using Deep Autoencoder Gaussian Mixture Model,'' \textit{Proc. IEEE VTC 2025-Fall}, 2025, pp. 1--5.
\end{thebibliography}
\end{document}